\documentstyle[amsfonts,aps]{revtex}
\begin{document}
\draft
\tighten
\twocolumn[\hsize\textwidth\columnwidth\hsize\csname   
@twocolumnfalse\endcsname                             
\title{ Diffusion-controlled annihilation $A + B \rightarrow 0$: The growth
of an $A$ particle island from a localized $A$-source in the $B$ particle sea}
\author{ Boris ~M.~Shipilevsky}
\address{ Institute of Solid State Physics, Chernogolovka,
Moscow district, 142432, Russia}
\date{\today}
\maketitle
\begin{abstract}
We present the growth dynamics of an island of particles $A$ injected from 
a localized $A$-source into the sea of particles $B$ and dying in
the course of diffusion-controlled annihilation $A+B\to 0$. We show that in 
the $1d$ case the island unlimitedly grows at any source strength $\Lambda$, 
and the dynamics of its growth {\it does not depend} asymptotically on the
diffusivity of $B$ particles. In the $3d$ case the island grows only at 
$\Lambda > \Lambda_{c}$, achieving asymptotically a stationary state 
({\it static island}). In the marginal $2d$ case the island unlimitedly 
grows at any $\Lambda$ but at $\Lambda < \Lambda_{*}$ the time of its 
formation becomes exponentially large. For all the cases the numbers of 
surviving and dying $A$ particles are calculated, and the scaling of the 
reaction zone is derived.

\end{abstract}
\pacs{82.20.-w, 05.40.-a}
]                                                 

\narrowtext

For the last two decades the reaction-diffusion system $A + B \rightarrow 0$,
where unlike species $A$ and $B$ diffuse and irreversibly react in a $d$
dimensional medium, has acquired the status of one of the most popular
objects. This attractively simple system, depending on the initial conditions,
displays a rich variety of phenomena and, depending on the interpretation of
$A$ and $B$ (chemical reagents, quasiparticles, topological defects, etc.),
it provides a model for a broad spectrum of problems \cite{rev}. A crucial
feature of many such problems is the dynamical {\it reaction front} - a
localized reaction zone which propagates between domains of unlike species. 
The simplest model of a reaction front, introduced more than a decade ago by 
Galfi and Racz [2], is a quasi-one-dimensional model for two initially 
separated reactants uniformly distributed on the left side ($x<0$) and on 
the right side ($x>0$) of the initial boundary. Taking the reaction rate in
the mean-field form $R(x,t)=ka(x,t)b(x,t)$($k$ being the reaction constant) 
Galfi and Racz discovered that in the long-time limit $kt\to\infty$
the reaction profile $R(x,t)$ acquires the universal scaling form
\begin{eqnarray}
R=R_{f}{\cal Q}\left(\frac{x-x_{f}}{w}\right),
\end{eqnarray}
where $x_{f}\propto t^{1/2}$ denotes the position of the reaction
zone center, $R_{f}\propto t^{-\beta}$ is the height, and
$w\propto t^{\alpha}$ is the width of the reaction zone. Since then, 
much work has been devoted to studying this problem
by different approaches. It has been shown [3 - 7] that the mean-field
approximation can be adopted at $d \geq d_{c}=2$ (with logarithmic
corrections in the $2d$ case), whereas in $1d$ systems fluctuations play the
dominant role and the explicit form of $R$ remains unknown. Nevertheless, the
scaling law (1) takes place in all dimensions with $\alpha=1/6$ at
$d \geq d_{c}$ and $\alpha=1/4$ at $d=1$, so that at any $d$ the system
demonstrates a remarkable property: on the diffusion length scale
$L_{D} \propto t^{1/2}$ the width of the reaction front asymptotically
unlimitedly contracts: $w/L_{D} \to 0$  as $t\to\infty$. Based on this 
property a general concept of the front dynamics, the quasistatic 
approximation (QSA), has been developed [3, 5, 8, 9]. The QSA consists
in the assumption that for sufficiently long times the dynamics of the front
is governed by two characteristic time scales. One time scale,
$t_{J}=-(d\ln J/dt)^{-1}$, controls the rate of change in the diffusive
current $J=J_{A}=J_{B}$ of particles arriving at the reaction zone.
The second time scale, $t_{f}\propto w^{2}/D$, is the equilibration time of
the reaction front. Assuming $t_{f}/t_{J}\ll 1$ from the QSA in the
mean-field case with equal species diffusivities $D_{A,B}=D$ it follows [3,8]
\begin{eqnarray}
R_{f}\sim J/w, \quad w\sim (D^{2}/Jk)^{1/3},
\end{eqnarray}
whereas in the $1d$ case $w$ acquires the $k$-independent form
$w\sim (D/J)^{1/2}$ [3,5,9]. The most important feature of the QSA is that
$w$ and $R_{f}$ depend on $t$ only through the boundary 
current $J(t)$, which can be calculated analytically without knowing the
concrete form of ${\cal Q}$. On the basis of the QSA the general description 
of the system $A+B\to 0$ with initially separated reactants have been 
obtained for arbitrary diffusion coefficients [10]. These results are in full 
agreement with extended numerical calculations and have been generalized 
recently to the case of nonmonotonic front motion [11].

The purpose of present paper is to apply the QSA to the long-standing
problem of growth of an $A$-particle island from a localized $A$-source
in the uniform $B$ particle sea. This important problem was first analyzed 
by Larralde {\it et al.} [12] for the special case of {\it static} sea 
($D_{B}=0$). Assuming that diffusing $A$ particles are injected at a single 
point into a reactive $d$-dimensional substrate $B$ and 
instantaneously react with $B$ upon contact, Larralde {\it et. al.} 
have studied the growth dynamics of the reacted region radius, $r_{f}(t)$, 
and the number of dying and that of surviving $A$ particles. 
Considering the reaction front 
dynamics as a Stefan problem, they have, in particular, shown that at any 
source strength $r_{f}$ asymptotically grows by the laws 
$\propto (t\ln t)^{1/2}, \propto t^{1/2}$ and $\propto t^{1/3}$ at $d=$1,2 
and 3, respectively. Subsequently, those results were generalized to the 
cases of imperfect reaction [13] and diffusion with a bias [14], however, 
as in [12], the $B$ particles were always presumed {\it "frozen"}. 
In this Rapid Communication we present a theory of growth of a 
$d$-dimensional $A$-island for the physically most important situation when 
{\it both $A$ and $B$ particles are mobile}. 
In the framework of the QSA we first consider the simplest "standard" 
case with equal species diffusivities, and then we extend the obtained 
results to the case of arbitrary nonzero diffusivities, thus revealing 
a rich general picture of the island growth for $d$=1, 2, 3. 

Let at $t\geq 0$ particles $A$ be injected with a rate $\Lambda$ at the 
point $\vec {r}=0$ of the uniform $d$-dimensional 
sea of particles $B$, distributed with a density $\rho$. Particles $A$ and 
$B$ diffuse with nonzero diffusion constants $D_{A,B}$ and upon contact with 
some nonzero probability annihilate, $A+B\to 0$. In the continuum version 
such process can be described by the reaction-diffusion equations
\begin{eqnarray}
\frac{\partial a}{\partial t} = D_{A}\nabla^{2} a - R + 
\Lambda\delta(\vec{r}), \quad 
\frac{\partial b}{\partial t} = D_{B}\nabla^{2} b - R
\end{eqnarray}
with the initial conditions $a(r,0)=0$, $b(r,0)=\rho$, and the boundary
condition $b(r\to \infty, t)=\rho$. Here $a(r,t)$ and $b(r,t)$ are the
mean local concentrations of $A's$ and $B's$ which, by symmetry, we assume 
to be dependent only on the radius, and $R(r,t)$ is the macroscopic reaction 
rate. 
{\it Equal species diffusivities}.--- To simplify the problem essentially we 
will first assume, as usually, $D_{A}=D_{B}=D$. 
The initial density of the sea, $\rho$, defines a natural scale 
of concentrations and a characteristic length scale of the problem - 
average interparticle distance $\ell=\rho^{-1/d}$. So, by measuring the    
length, time and concentration in units of $\ell, \ell^{2}/D$ and $\rho$, 
respectively, we introduce the dimensionless source strength
$\lambda =\Lambda\ell^{2}/D$ 
and the dimensionless reaction constant $\kappa=k/\ell^{(d-2)}D$.
Defining then the difference 
concentration $s(r,t)=a(r,t)-b(r,t)$ we come from (3) to the simple 
diffusion equation with source
\begin{eqnarray}
\partial s/\partial t = \nabla ^{2} s + \lambda \delta(\vec{r})
\end{eqnarray}
at the initial and boundary conditions $s(r,0)= s(r\to \infty, t)= -1$.
According to (4) in the course of injection in the vicinity of the source 
there arises a region of $A$ particle excess, $s>0$, which expands with time. 
The central idea of the paper is that, by analogy with the Galfy-Racz 
problem, a narrow reaction front has to form at this region boundary, 
of which the law of motion, $r_{f}(t)$, according to QSA, can be derived 
from the remarkably simple condition $s(r_{f},t)=0$. Then, under the 
assumption that on the scale $r_{f}$ the front width $w$ can be neglected 
$w/r_{f}\ll 1$, i.e., setting that $a=s,b=0$ at $r < r_{f}$ whereas $a=0, 
b=|s|$ at $r > r_{f}$, the number of surviving $N_{A}(t)$ and that of dying 
$N_{\times}(t)$ $A$ particles is immediately derived from the condition   
\begin{eqnarray}
N_{A}=\lambda t- N_{\times}=
\Omega_{d}\int_{0}^{r_{f}}s(r,t)r^{d-1}dr
\end{eqnarray}
with $\Omega_{1}=2, \Omega_{2}=2\pi, \Omega_{3}=4\pi$. 
By calculating in the limit $w/r_{f} \ll 1$ the current of particles 
in the vicinity of the front, $J=-\partial s/\partial r|_{r=r_{f}}$, 
from Eqs.(2) one can easily obtain the law $w(t)$ and define, in the end, 
a self-consistent condition of crossover to a quasistatic scaling regime (1). 
We start with an analysis of the behavior of $r_{f}(t), N_{A}(t)$ and 
$N_{\times}(t)$ for each dimension separately.
 
{\it 1d case.}---In $1d$ the solution to Eq.(4) has the form 
\begin{eqnarray}
s(r, t) = (\sqrt {\lambda^{2}t}){\rm ierfc}(r/2\sqrt {t}) - 1,
\end{eqnarray}
whence, according to condition $s(r_{f}, t)=0$, the equation of motion of
the reaction front center, $r_{f}(t)$, is
\begin{eqnarray}
{\rm ierfc} 
(r_{f}/2\sqrt {t}) = 1/\sqrt {\lambda^{2} t},
\end{eqnarray}
where ${\rm ierfc}(\zeta)= \int_{\zeta}^{\infty}{\rm erfc}(v)d v=
e^{-\zeta^{2}}/\sqrt {\pi} - \zeta {\rm erfc}(\zeta)$. From (6),(7) 
it formally follows that an excess of $A$ particles forms in a time 
$t_{c}=\pi/\lambda^{2}$. It is, however, clear that a continual 
approximation comes into play at times $t\gg {\rm max}(1, 1/\lambda)$, 
therefore at early island formation stages one can distinguish two 
qualitatively different island growth regimes: i) $\lambda \ll 1$, 
when the island formation proceeds under conditions of death of the 
majority of injected particles and ii) $\lambda \gg 1$, when 
the island forms long before the beginning of intensive annihilation.
Let us consider first the limit $\lambda \ll 1$. In this limit an interval 
between injection acts, $\delta t_{\lambda}=1/\lambda$, is quite large, 
therefore in the case of perfect reaction each injected particle dies 
long before the next one appears until the distance to the nearest sea 
particle, $\sim \lambda t - \sqrt{t}$, becomes comparable with the 
characteristic diffusion length $\sqrt{\delta t_{\lambda}}$. Whence 
for the time of beginning of the injected particles accumulation we find 
$t_{b}\propto t_{c}\propto \lambda^{-2}$, which reveals the sense of 
$t_{c}$. Assuming $\epsilon=(t-t_{c})/t_{c}\ll 1$ we have 
$\zeta_{f}=r_{f}/2\sqrt{t} \ll 1$ and hence from (5)-(7) we find 
$$
r_{f}\sim \epsilon/\lambda, \quad 
N_{A}\sim \epsilon^{2}/2\lambda, \quad 
N_{\times}\sim \pi(1+\epsilon)/\lambda,
$$
whence it follows $N_{A}/N_{\times}\propto \epsilon^{2}$.
Calculating from (6),(7) the current  
$J\sim\lambda(1-\epsilon\lambda/\pi)/2$ and assuming $\kappa$ not to be 
too small [9]($\kappa > \sqrt{\lambda}$) from the fluctuation law 
$w\sim J^{-1/2}\sim 1/\sqrt{\lambda}$ we find 
$w/r_{f}\sim \sqrt{\lambda}/\epsilon \sim 1/\sqrt{N_{A}}$ 
and $t_{f}/t_{J}\propto \lambda^{2}$. Thus, the condition of crossover to 
the regime of quasistatic front is $\epsilon \gg \sqrt{\lambda}
(N_{A}\gg 1)$. Defining a minimal island by condition 
$w/r_{f}\sim N_{A}\sim 1$ for the island formation time we have  
$\epsilon_{b}\sim \sqrt{\lambda}$ and hence 
$t_{b}\sim t_{c}(1+\sqrt{\lambda})$ 
and $r_{b}\sim 1/\sqrt{\lambda}$. In the long-time limit 
${\cal T}=t/t_{c}\gg 1$ from (7) it follows $\zeta_{f}\gg 1$ and we 
can rewrite (7) in the form 
$2e^{\zeta_{f}^{2}}\zeta_{f}^{2}=\sqrt {\cal T}$ whence we 
obtain {\it exact} asymptotics 
\begin{eqnarray}
r_{f}= \sqrt {2t \ln{\cal T}}
(1-\ln\ln{\cal T}/\ln{\cal T}+ \cdots),
\end{eqnarray}
and from (5),(6), and (8) we find 
$N_{A}=\lambda t[1- O(\sqrt {\ln{\cal T}/{\cal T}})], 
N_{\times}=r_{f}(2+\zeta_{f}^{-2}+\cdots)=\sqrt {8t\ln{\cal T}}$.
Consider now the limit $\lambda \gg 1$. It is evident that in this 
limit a multiparticle "cloud" forms long before the beginning of noticeable 
annihilation, therefore the stage of developed reaction (8), 
$t\gg 1(\gg t_{c})$ is preceded here by the stage of purely diffusive 
expansion of the cloud, $1/\lambda\ll t \ll 1$. The statistical theory of 
diffusive cloud expansion has been developed recently in the work [15]. 
According to [15] in the "collective" regime ($\lambda t \gg 1$) the 
"radius" of an $1d$ cloud grows (in our units) by the law 
$r_{+}=\sqrt {4t\ln(\lambda t)}$. Remarkably, $r_{+}$ and $r_{f}$ "join" 
exactly in the front formation zone $(t\sim 1, r_{+}\sim r_{f})$ whereas at 
$t\gg 1$$r_{f}$ begins to grow slower than $r_{+}$, as it has to. So, forming 
in qualitatively different regimes from 
$\wp=N_{\times}/N_{A}\gg 1 (\lambda \ll 1)$ to 
$\wp\ll 1 (\lambda\gg 1)$ the $1d$ island at any $\lambda$ crosses over to 
the universal growth regime (8) with an unlimited decay of the dying 
particles fraction $\wp\propto \sqrt{\ln{\cal T}/{\cal T}}\to 0$.
It remains for us to reveal conditions of quasistaticity of the front (8). 
According to (6) in the limit $t,{\cal T}\gg 1$ current $J\sim \sqrt 
{\ln{\cal T}/t}$. Thus assuming $\kappa$ not to be too small 
$(\kappa >\sqrt {J})$ we find $w\sim J^{-1/2}\sim (t/\ln{\cal T})^{1/4}$ 
whence $w/r_{f}\sim (t\ln^{3}{\cal T})^{-1/4}$ and $t_{f}/t_{J}\sim
(t\ln{\cal T})^{-1/2}$. As $\zeta_{f}\gg 1$, the conditions $w/r_{f}, 
t_{f}/t_{J}\ll 1$ ought to be supplemented by a more strict requirement 
of equality of currents at both front sides 
$w\ll {\cal L}=-(\frac{d\ln J}{dr})^{-1}\mid_{r=r_{f}}=r_{f}/2\zeta^{2}_{f}$.
Calculating $w/{\cal L}\sim (\ln{\cal T}/t)^{1/4}$ we arrive at the 
requirement $t\gg {\rm max}(t_{c},\ln{\cal T})$.  

{\it 2d case.}---In $2d$ the solution to Eq.(4) has the form
\begin{eqnarray}
s(r, t) = -(\lambda/4\pi){\rm Ei}(-r^{2}/4t) - 1,
\end{eqnarray}
whence, according to condition $s(r_{f}, t)=0$, the equation 
of motion of the reaction front center, $r_{f}(t)$, is
\begin{eqnarray}
{\rm Ei}(-r_{f}^{2}/4t) = -4\pi/\lambda.
\end{eqnarray}
Here ${\rm Ei}(-\zeta)=-\int_{\zeta}^{\infty}d v e^{-v}/v$ is the 
exponential integral which has the asymptotics
${\rm Ei}(-\zeta)=\ln(\gamma \zeta) + \cdots$ at $\zeta \ll 1$
($\gamma= 1.781...$) and
$-e^{-\zeta}/\zeta + \cdots$ at $\zeta \gg 1$.
From (10) it follows that $r_{f}$ grows by the law
\begin{eqnarray}
r_{f}=2\sqrt {\alpha t},
\end{eqnarray}
where $\alpha$ is the root of the equation 
${\rm Ei}(-\alpha) = -\lambda_{*}/\lambda, \lambda_{*}=4\pi$
and has the asymptotics 
$\alpha = e^{-\lambda_{*}/\lambda}/\gamma$ at $\lambda \ll \lambda_{*}$ 
and $\alpha=\ln(\lambda/\lambda_{*}\alpha)$ at $\lambda \gg \lambda_{*}$.
From (5), (9), (11) it follows
\begin{eqnarray}
N_{A}=\lambda t(1-e^{-\alpha}), \quad
N_{\times}=\lambda te^{-\alpha}.
\end{eqnarray}
We conclude that in $2d$ the island growth rate $\alpha$ and the 
dying-to-surviving particles ratio $\wp$ {\it do not vary in time}: 
at large $\lambda \gg \lambda_{*}$ the majority of particles survive, 
$\wp\sim \ln\lambda/\lambda$, whereas at small 
$\lambda \ll \lambda_{*}$ the majority of particles die, 
$\wp\sim e^{\lambda_{*}/\lambda}$. The most interesting consequence of (10) 
consists in the exponentially strong decrease of the growth rate  
in the region $\lambda < 1$. Defining a minimal island through 
the condition $N_{A}\sim 1$ for its formation time at $\lambda < 1$ 
we have $t_{b}\sim e^{\lambda_{*}/\lambda}/\lambda$ whence it is seen 
that at $\lambda\ll \lambda_{*}$ the island growth is actually  
{\it suppresed}. Calculating current 
$J=\lambda/\lambda_{*}e^{\alpha}\sqrt{\alpha t}$ for the scaling of 
the reaction zone from (2) we find $w\sim (t/t_{w})^{1/6}, 
t_{w}=(\kappa\lambda/e^{\alpha}\sqrt{\alpha})^{2}$. At $\lambda \ll 
\lambda_{*}$ this yields $\sqrt{t_{f}/t_{J}}\ll w/r_{f}\sim 
(t_{b}/\kappa t)^{1/3}$ whereas at $\lambda \gg \lambda_{*}$ 
we have $\sqrt{t_{f}/t_{J}}\ll w/{\cal L}\sim (\ln\lambda/\kappa t)^{1/3}$.
Thus, crossover to the quasistatic regime occurs at times 
$\kappa t\gg t_{b}$ and $\kappa t\gg \ln\lambda$, respectively 
(note that $\kappa < 1$, being $\sim 1$ for perfect reaction).  

{\it 3d case.}---In $3d$ the solution to Eq.(4) has the form
\begin{eqnarray}
s(r, t) = (\lambda/4\pi r){\rm erfc}(r/2\sqrt {t}) - 1,
\end{eqnarray}
whence, according to condition $s(r_{f}, t)=0$, the equation of motion of the 
reaction front center, $r_{f}(t)$, is 
\begin{eqnarray}
{\rm erfc}(r_{f}/2\sqrt {t}) = 4\pi r_{f}/\lambda.
\end{eqnarray}
From (14) it follows that $\zeta_{f}(t)$ decreases indefinitely so that 
at large $t/t_{s}\gg 1$ the front radius, by the law  
$r_{f}=r_{s}[1-O(\sqrt {t_{s}/t})]$ with the characteristic time 
$t_{s}=(\lambda/\lambda_{*})^{2}$,  reaches a stationary value 
\begin{eqnarray}
r_{f}(t/t_{s}\to\infty)=r_{s}=\lambda/\lambda_{*}.
\end{eqnarray}
According to (5), (13), (14) in this limit 
$N_{A}=\frac {2\pi}{3}r_{s}^{3}[1- O(\sqrt {t_{s}/t})], 
N_{\times}=\lambda t[1- O(t_{s}/t)]$ therefore, in contrast to the 
$1d$ case, at any $\lambda$ all the injected particles die.
The steady-state current $J_{s}=\lambda_{*}/\lambda$, whence,  
according to (2) $w_{s}\sim (\lambda/\lambda_{*}\kappa)^{1/3}$ 
and $w_{s}/r_{s}\sim(\lambda_{*}/\lambda \sqrt {\kappa})^{2/3}$.
Defining a minimal stationary island through the condition 
$w_{s}/r_{s}\sim 1$ we conclude that in the $3d$ case the island forms 
only when the injection rate exceeds a {\it critical} value 
$\lambda_{c}\sim \lambda_{*}/\sqrt {\kappa}$.
A maximal value of $\kappa$, attainable in the perfect reaction 
limit, is $\kappa_{p}\sim \sigma/\ell$ ($\sigma$ being the size of 
particles), therefore $\lambda_{c} > \lambda_{*} \gg 1$. 
By rewriting Eq.(14) in the form 
${\rm erfc}(\zeta_{f})/\zeta_{f}= 2\sqrt {t/t_{s}}$ one can easily see 
that at high injection rates $\lambda/\lambda_{*}\gg 1$ the
stationary stage ($t\gg t_{s}$) is preceded by an intermediate stage 
$1\ll t \ll t_{s}$ wherein the island grows by the law
\begin{eqnarray}
r_{f}=\sqrt {2t \ln(t_{s}/t)}(1-\ln(\sqrt{\pi}\omega)/\omega+\cdots),
\end{eqnarray}
where $\omega= \ln(t_{s}/t)$. According to (5), (13), (16) 
at this stage $N_{A}=\lambda t[1-O(\omega^{3/2}\sqrt {t/t_{s}})], 
N_{\times}=\frac {4\pi}{3}r_{f}^{3}[1+ O(\omega^{-1})]$ and, therefore 
the majority of particles have yet survived, 
$\wp \sim \omega^{3/2}\sqrt {t/t_{s}}\ll 1$. 
Calculating the current $J\sim \sqrt {\omega/t}$ we find from (2) 
$w\sim(t/\kappa^{2}\omega)^{1/6}$ whence 
$\sqrt{t_{f}/t_{J}}\ll w/{\cal L}\sim (\omega/\kappa t)^{1/3}$. 
Thus, the formed front condition reads $t\gg \ln(t_{s}/t)/\kappa$. 
According to [15] the radius of a $3d$ 
cloud, which expands in the absence of reaction, in our units has the 
form $r_{+}=\sqrt {2t\ln[(\lambda\sigma/\ell)^{2}t/4\pi]}$. Comparing 
$r_{+}$ and $r_{f}$ suggests that $r_{f}$ begins to lag behind $r_{+}$ 
at times $t\gg\ell/\sigma \sim 1/\kappa_{p}$ in remarkable 
agreement with the above estimation. 

To sum up the above, as key results we distinguish 
a){\it self-consistent analytic description} of the island growth 
since the moment of its formation and b) {\it anomalously slow} 
island growth at $\lambda < \lambda_{*}$ in $2d$ and 
{\it complete suppression} of its growth at $\lambda < \lambda_{c}$ in 
$3d$, which contrast sharply with growth asymptotics in the static sea [12].

{\it Arbitrary ratio of diffusivities}.--- Let us now extend the 
analysis to a general case $0<{\cal D}= D_{B}/D_{A}<\infty$, 
nondimensionality with respect to $D=D_{A}$ being retained. 
We present final results here, a detailed discussion of which 
will be given elsewhere.
 
[d=1]. Comparing our results with the results of Ref.[12]
we find that long-$t$ asymptotics (LTA)(8) for ${\cal D}=1$ converges to the 
LTA for ${\cal D}=0$. We thus conclude that the growth 
of the $1d$ island {\it does not depend} asymptotically on the diffusivity of 
$B$ particles. This conclusion is the consequence of the evident fact that at
$r_{f}/\sqrt {{\cal D}t} \gg 1$ particles $B$ can be regarded as effectively 
static, so as $r_{f}/\sqrt {t}\propto\sqrt {\ln t}\to\infty$
the LTA of the island growth at any $0< {\cal D}< \infty$ must converge 
to the LTA for static sea. In the interval $0<{\cal D}< 1$ 
the time of crossover to the LTA does not alter appreciably although the time 
of $A$ particle accumulation, $t_{b}$, shifts considerably at small 
$\lambda$(comparing $\lambda t -\sqrt{{\cal D}t}$ with 
$\sqrt {\delta t_{\lambda}}$ we find that in the interval 
${\cal D}\ll \sqrt {\lambda}\ll 1$ the value of 
$t_{b}\propto \lambda^{-3/2}$ does not depend on ${\cal D}$ whereas 
at $\sqrt {\lambda}\ll {\cal D}$ it grows with ${\cal D}$ by the 
law $t_{b}\propto {\cal D}/\lambda^{2}$). At ${\cal D} \gg 1$ the time 
of crossover to the LTA becomes {\it exponentially large}, 
$t\gg e^{\cal D}/\lambda^{2}$, therefore the transient dynamics in this 
limit requires to be specially considered (note that for this case at 
small $\sqrt {\lambda/{\cal D}}\ll 1$ we again find 
$t_{b}\propto {\cal D}/\lambda^{2}$).
  
[d=2]. In the $2d$ case the solution of the problem with a source possesses 
a remarkable property: $s(r,t)=f(\frac {r}{\sqrt {t}})$. Using this property 
and assuming $w/r_{f}(t\to\infty)\to 0$ it is easy to check that the 
solution of Eqs.(3) has to read 
$a= -(\frac {\lambda}{4\pi}){\rm Ei}(-\zeta^{2})-{\cal A}, b=0$ at
$r<r_{f}$ and $a=0, b=1 +{\cal B}{\rm Ei}(-\zeta^{2}/{\cal D})$ at 
$r>r_{f}$ with $\zeta=\frac {r}{2\sqrt {t}}({\cal A},{\cal B}=
{\rm const})$. Equalizing $A's$ and $B's$ currents at $r=r_{f}$ we come 
to the laws (11),(12) with the {\it exact} equation for $\alpha$ at arbitrary 
${\cal D}$:  
\begin{eqnarray}
{\rm Ei}(-\alpha/{\cal D})=
-(4\pi{\cal D}/\lambda)e^{\alpha({\cal D}-1)/{\cal D}}.
\end{eqnarray}
In the limit ${\cal D}\ll 1$ from (17) it follows
\begin{eqnarray}
\alpha\sim
\left\{\begin{array}{ll}
{\cal D}e^{-\lambda_{*}{\cal D}/\lambda}/\gamma, \quad
&\lambda/\lambda_{*}\ll{\cal D} \nonumber\\
\lambda/\lambda_{*}, \quad &{\cal D}\ll\lambda/\lambda_{*}\ll 1 \nonumber\\
\ln(\lambda/\lambda_{*}\alpha), \quad &\lambda/\lambda_{*}\gg 1. \nonumber\\
\end{array}
\right.
\end{eqnarray}
In the opposite limit ${\cal D}\gg 1$ from (17) it follows 
\begin{eqnarray}
\alpha\sim
\left\{\begin{array}{ll}
{\cal D}e^{-\lambda_{*}/\tilde{\lambda}}/\gamma, \quad
&\tilde{\lambda}/\lambda_{*}\ll\ln^{-1}{\cal D} \nonumber\\
\ln[\tilde{\lambda}\ln({\cal D}/\alpha)/\lambda_{*}], \quad
&\ln^{-1}{\cal D}\ll\tilde{\lambda}/\lambda_{*}\ll e^{\cal D} 
\nonumber\\
\ln(\lambda/\lambda_{*}\alpha), \quad &\tilde{\lambda}/\lambda_{*}\gg
e^{\cal D},
\nonumber\\
\end{array}
\right.
\end{eqnarray}
where $\tilde{\lambda}=\lambda/{\cal D}$. Thus, at strong difference 
of diffusivities we find three characteristic growth regimes: 
$\alpha\ll \alpha_{-}={\rm min}(1,{\cal D})$(I), 
$\alpha_{-}\ll\alpha\ll\alpha_{+}$(II) and 
$\alpha\gg \alpha_{+}={\rm max}(1,{\cal D})$(III). In regime I the majority 
of particles die, and the island growth 
{\it does not depend} on $D_{A}$. In regime III the majority of particles 
survive, and the island growth {\it does not depend} on $D_{B}$. 
In intermediate regime II at ${\cal D}\ll 1$ the majority of particles 
die and the island growth does not depend on $D_{B}$ whereas at 
${\cal D}\gg 1$ the majority of particles survive and the island 
growth depends on the both diffusivities.  
Essentially, that at any finite ${\cal D}$, as $\lambda$ decreases, 
the island growth crosses over to the regime of 
"exponential suppression" I which disappears only in the limit 
${\cal D}\to 0$.

[d=3]. One can easily check that in the general case ${\cal D}\neq 0$, 
like in the case ${\cal D}=1$, the $3d$ island has to reach asymptotically 
a stationary state. Indeed, assuming $\frac {w_{s}}{r_{s}}
\to 0$ in the steady state limit $t\to\infty$ from Eqs.(3) we find 
$a_{s}=\Theta(r_{s}-r){\cal D}(\frac {r_{s}}{r} - 1), 
b_{s}=\Theta(r-r_{s})(1-\frac {r_{s}}{r})$, where $\Theta(x)$ 
is the Heaviside step function and stationary radius 
$$
r_{s}=\lambda/\lambda_{*}{\cal D}=\Lambda \ell^{2}/4\pi D_{B}
$$
{\it does not depend} on $D_{A}$. For the particle number we find 
$N_{A}^{s}= \frac {2\pi}{3}{\cal D}r_{s}^{3} \sim
(\lambda/\lambda_{*})^{3}/{\cal D}^{2}$ whence we obtain the lower  
island formation boundary $\lambda/\lambda_{*}\gg {\cal D}^{2/3}$ and 
conclude that a mean island density $<a>_{s}={\cal D}/2$ 
{\it does not depend} on $\lambda$: 
the island is always concentrated (with respect to the sea) 
at ${\cal D} \gg 1$ and is always rarefied at ${\cal D} \ll 1$. According to 
[10] for ${\cal D}\neq 1$ $w_{s}\sim (\frac {{\cal D}}{\kappa J})^{1/3}
\sim (\frac {r_{s}}{\kappa})^{1/3}$ whence it follows 
$w_{s}/r_{s}\sim (r_{s}^{2}\kappa)^{-1/3}$. Thus the necessary condition for 
the $3d$ island formation reads
$$
\lambda \gg \lambda_{c}\sim (\lambda_{*}{\cal D}/\sqrt {\kappa})
{\rm max}(1, \sqrt {\kappa/{\cal D}^{2/3}}).
$$
We have been unable to describe analytically the complete kinetics 
of crossover to the steady state for arbitrary ${\cal D}\neq 1$. However, 
in the interval $0< {\cal D}< 1$ intermediate asymptotics of the $3d$ island 
growth can be revealed based on simple arguments. Indeed, at ${\cal D}<1$ 
in the limit $\sqrt {{\cal D}t}\gg r_{f}$ we have $r_{f}\sim r_{s}$.
In the opposite limit $\sqrt {{\cal D}t}\ll r_{f}$ the sea is effectively 
static, therefore the island ought to grow by the law 
$r_{f}\sim (\lambda t)^{1/3}$[12]. From the both conditions for the 
crossover time it follows $t_{s}\propto \lambda^{2}/{\cal D}^{3}$ 
so that at $t< t_{s}$ the particle number grows by the law 
$N_{A}\sim (\lambda^{5/2}t)^{2/3}$, and the density decays to 
$<a>_{s}$ by the law $<a>\sim (\lambda^{2}/t)^{1/3}$. Clear that 
at $\lambda \gg 1$ this asymptotics is preceded by the law (16) 
(crossover time $t_{i}\sim \lambda^{2}$) whereas at 
$\lambda_{c} \ll \lambda \ll 1$ the island formation time  
$t_{b}\propto \lambda^{-5/2}$. In the limit ${\cal D}\to 0$ 
it follows $\lambda_{c}\to 0, r_{s}\to \infty, 
t_{s}\to \infty$ and we are coming back to the unlimited island 
growth at arbitrary finite $\lambda$.  

This research was financially supported by the RFBR through Grant
No. 02-03-33122.

\end{document}